\documentstyle[12pt,epsfig]{article}
\newcommand{\beq}{\begin{equation}}
\newcommand{\eeq}{\end{equation}}
\newcommand{\beqar}{\begin{eqnarray}}
\newcommand{\eeqar}{\end{eqnarray}}
\newcommand{\bfig}{\begin{figure}}
\newcommand{\efig}{\end{figure}}

\newcommand{\bd}{\begin{itemize}} 
\newcommand{\ed}{\end{itemize}} 
\newcommand{\bc}{\begin{center}}
\newcommand{\ec}{\end{center}}
\newcommand{\be}{\begin{equation}}
\newcommand{\ee}{\end{equation}}

\newcommand{\ba}{\begin{array}}
\newcommand{\ea}{\end{array}}

\begin{document}
\begin{center}
\Large{Boson expansion methods applied to a two-level model in the
  study of multiple giant resonances}

\vspace{1.cm}

\large{C. Volpe} \\
Groupe de Physique Th\'eorique, Institut de Physique Nucl\'eaire,\\
F-91406 Orsay Cedex, France\\ 
\vspace{0.5cm}
\large{Ph. Chomaz}\\
GANIL, B.P. 5027,\\ 
F-14076 Caen Cedex 5, France\\
\vspace{0.5cm}
\large{M.V. Andr\'{e}s }\\
Departamento de F\'{\i}sica At\'omica, Molecular y
Nuclear,\\ Universidad de Sevilla, Apdo 1065, 41080 Sevilla, Spain\\
\vspace{0.5cm}
\large{F. Catara}\\
Dipartimento di Fisica dell'Universit\`a and 
INFN, Sezione di Catania,\\
 I-95129 Catania, Italy\\                                     
\vspace{0.5cm}
\large{E.G. Lanza}\\
Dipartimento di Fisica dell'Universit\`a and 
INFN, Sezione di Catania,\\
 I-95129 Catania, Italy\\                                     
and\\
Departamento de F\'{\i}sica At\'omica, Molecular y
Nuclear,\\ Universidad de Sevilla, Apdo 1065, 41080 Sevilla, Spain\\

\end{center}

\vspace{0.5cm}
\begin{abstract}
We apply boson expansion methods to an extended Lipkin-Meshkov-Glick
model including anharmonicities in analogy with previous microscopic
calculations.  We study the effects of different approximations
present in these calculations, among which the truncation of the
hamiltonian and of the space, in connection with the study of the
properties of two-phonon and three-phonon states. By comparing the
approximate results on the spectrum with the exact ones we conclude
that the approximations made in the microscopic calculations on two-phonon
states are well justified. We find also that a good agreement with the
exact results for the three-phonon state is obtained by using a bosonic 
hamiltonian truncated at the fourth order. This result makes us confident
that such approximation can be used in realistic calculations, thus allowing
a theoretical study of triple excitations of giant resonances.
\end{abstract}

\section{Introduction}

Collective excitations have been known for many years in nuclear
physics \cite{BM}, both in the low-lying spectra and in the Giant
Resonance (GR) region. A basic microscopic theory for such collective
modes is the Random Phase Approximation (RPA) \cite{RO,RS} which can
be seen as the lowest order in a boson expansion such that the
hamiltonian can be put in the form of a sum of hamiltonians of
harmonic oscillators, each one corresponding to a collective mode
(phonon). Therefore, RPA predicts the existence of one-phonon,
two-phonon,...etc states with a harmonic spectrum. In addition to the
well known low-lying two-phonon states, recently heavy ion inelastic
scattering experiments at intermediate and relativistic energies and
double charge-exchange reactions have shown the existence of states in
the high excitation energy region which can be described as a GR built
on top of another GR \cite{CF,EM}. The study of the properties of
these states allows to test our comprehension of the GR's as small
amplitude vibrations and therefore the harmonic picture. The
systematics on the energies and widths is in qualitative agreement
with the harmonic approximation, namely the energy of a doubly excited
GR is nearly twice that of the single GR and its width is between
$\sqrt 2$ and 2 times larger. However, the inelastic cross sections,
when calculated within the harmonic picture, are almost always smaller
than the measured ones. In particular, the data concerning Coulomb
excitation show a discrepancy ranging from $30 \%$ up to a factor 4,
according to the nucleus studied. In order to understand the origin of
this discrepancy, corrections to the harmonic approximation have been
proposed \cite{VCCAL} by including anharmonicities in the internal
hamiltonian and non-linearities in the external field. As shown in
\cite{VCCAL}, small anharmonicities in the excitation spectrum of the
target nucleus can lead to a large enhancement of the Coulomb
excitation cross section. The model used there was an oversimplified
one, namely the target was described as an anharmonic linear
oscillator. The parameters of the cubic and the quartic terms in its
hamiltonian were fixed so that the energy of the second excited state
was shifted down by $\approx 2~MeV$ with respect to twice the energy
of the first excited state. In \cite{BD} a 3-dimensional extension of
this model was considered and similar effects coming from
anharmonicities were found. We remark that, in the case of Coulomb
excitation of $^{136}$Xe in the reaction $^{136}$Xe + $^{208}$Pb at
E/A = 700 MeV, the experimentally observed peak, interpreted as the
Double Giant Dipole Resonance (DGDR), is shifted by $\approx -2$~MeV
from that expected in the harmonic case \cite{Schmidt}.  A more
refined study was performed in \cite{LACCV}, where anharmonicities and
non-linearities were included by starting from RPA and extending it by
means of boson mapping techniques \cite{RS,CCG}. Such microscopic
approach was applied in \cite{LACCV} to realistic cases. In the case
of $^{208}$Pb + $^{208}$Pb at E/A = 641 MeV, for example, it was found
that the inclusion of anharmonicities and non-linearities gives rise
to an enhancement of $30 \%$ of the cross section in the region of the
DGDR, bringing the theoretical results closer to the experimental
ones. Important contributions coming from other two-phonon states in
the same energy region were also found.

\noindent The studies reported in the above quoted papers have of
course some limitations.  The model used in \cite{VCCAL} is clearly
very schematic and has, among others, the drawback that it is based on
a purely bosonic description so that Pauli blocking effects are not
included. The approach presented in \cite{LACCV} is certainly much
more realistic and the effects of the Pauli exclusion principle are
taken into account to some extent. However, for computational reasons,
only one- and two-phonon states were considered there.

\noindent
The purpose of this paper is to present an analysis based on an
extension of the Lipkin-Meshkov-Glick (LMG) model \cite{LMG} including
a residual interaction between the phonons, in analogy with the
microscopic approach used in ref.\cite{LACCV}.  This model has several
advantages with respect to the others.  Because of its group
structure, this model is exactly solvable. Besides, it directly takes
into account the Pauli principle.  In this context, we discuss several
approximations, corresponding to the cases considered in
\cite{VCCAL,LACCV} and, by direct comparison with the exact results,
we test how severe are the limitations of these approximations.  We
focus on one hand on two-phonons states, studied in ref.\cite{LACCV}
and on the other hand on the properties of three-phonon states.  In
fact, in order to have a more stringent test on the validity of the
harmonic picture, experimental and theoretical studies of triple
excitation of GR's should be envisaged.  From the theoretical point of
view, the same approach as the one in \cite{LACCV} can be used, but
this would imply huge calculations.  In order to make them feasible in
the context of boson expansion methods, one may consider an
approximation in which the same fourth order hamiltonian used in
\cite{LACCV} is diagonalized in a space containing up to 3-phonon
states.  We present a test on how well this approximation works, in
the context of the extended LMG hamiltonian.  We will see that the
exact results are very close to the approximate ones in all the cases
studied for both the second and the third excited states.

The paper is organized as follows. In section 2 the LMG model is
shortly reviewed and our extension introduced.  In section 3 the boson
mapping of this hamiltonian is presented.  In section 4 the
approximate results on the energy of the first, second and third
excited states, obtained by truncating at different levels the
mapping, are compared among themselves and with the exact
calculations.

\section{The model}
\subsection{The Lipkin-Meshkov-Glick model}
In the original Lipkin-Meshkov-Glick (LMG) model \cite{LMG} a finite
number $\Omega $ of particles can be arranged in two levels separated by
$\varepsilon $ in energy. $\Omega $ different quantum states are
available in each level.  Each particle is therefore identified by two
quantum numbers. The first $ \sigma $ indicates if the particle is
in the upper ($\sigma =+$) or in the lower ($\sigma =-$) energy
level. The second quantum number $s$ is related to the $\Omega $
different available quantum states in each level.  We will consider
systems whose hamiltonian can be written as function of the operators
$K_{+}$ , $K_{-}$ , $K_{0}$ :

\begin{eqnarray}
K_{+} &=&\sum_{s=1}^{\Omega }a_{+,s}^{\dagger }a_{-,s} \label{su2}\\ 
K_{-} &=&(K_{+})^{\dagger }=\sum_{s=1}^{\Omega }a_{-,s}^{\dagger }a_{+,s} 
\nonumber \\
K_{0} &=&{\frac{1}{2}}\sum_{s=1}^{\Omega }(a_{+,s}^{\dagger
}a_{+,s}-a_{-,s}^{\dagger }a_{-,s})        \nonumber 
\end{eqnarray}
where $a_{\sigma ,s}^{\dagger }$ (and the hermitian conjugate
$a_{\sigma ,s}$ ) creates (annihilates) a particle in the quantum
state $\left( \sigma ,s\right) $.  Since the operators $K_{+}$ ,
$K_{-}$ , $K_{0}$ satisfy the SU(2) commutation relations :

\begin{equation}
\lbrack K_{+},K_{-}]=2K_{0}\;\;\;\;\;\;\;\;\;
[K_{0},K_{\pm }]=\pm K_{\pm }  \label{com}
\end{equation}
they are often called the quasi-spin operators.  According to
(\ref{su2}), there can not be two particles in the same quantum state,
that is the Pauli exclusion principle is included in the model.  A
system of $\Omega$ particles has $2^{\Omega}$ states. However, since
the hamiltonian is a function of the generators of SU(2) and $K^{2}$
is a Casimir operator of such algebra, the space can be separated into
subspaces, each corresponding to an eigenvalue $K$. The state with all
particles in the lower level, $\sigma =-$, is an eigenvector of $K^{2}$
and $K_{0}$, belonging to their maximum and minimum eigenvalue,
respectively. This state, with the ones generated by applying $K_{+}$
to it, are the elements of a subspace of dimension $\Omega +1$, that
we will denote by $\left| \Omega /2,m\right\rangle $ with $m\in \left[
-\Omega /2,\Omega/2 \right]$.

\noindent Following (\ref{su2}) the original hamiltonian of LMG
\cite{LMG} can be written as :

\begin{equation}
H_{LMG}=\varepsilon \bar{K}_{0}+V_{1}K_{+}K_{-}+V_{2}(K_{+}K_{+}+K_{-}K_{-})
\label{LMG}
\end{equation}
where
\begin{equation}
\bar{K}_{0}=K_{0}+\Omega/2
\label{bar}
\end{equation}
The $\varepsilon \bar{K}_0$ term can be viewed as corresponding to the
Hartree-Fock part.  The lowest eigenstate of $K_0$ corresponds then to
the uncorrelated HF ground state $\left| \Omega/2,-\Omega/2
\right\rangle$. In order to stress the analogy with the microscopic
calculations in \cite{LACCV}, we will often denote by $h$ (hole) a
$(-,s)$ state, i.e. a single particle state which is occupied in
$\left| \Omega/2,-\Omega/2 \right\rangle$, and by $p$ (particle) a
$(+,s)$ single particle state.  Therefore, by looking at
eq.(\ref{su2}), we can say that the residual interaction in
(\ref{LMG}) only contains particle-hole terms
$a^{\dagger}_{p}a^{\dagger}_{p^{\prime}}a_ha_{h^{\prime }}$ and
$a^{\dagger}_{p}a^{\dagger}_ha_{p^{\prime}}a_{h^{\prime}}$ terms, that
is those usually included in RPA.

\noindent If $V_{1}=V_{2}=0$ then $H_{LMG}=\varepsilon \bar{K}_{0}$.
The corresponding eigenvalues are $n\varepsilon $ with $n\in \left[
0,\Omega \right] .$ From now on we will denote these states by
$|n>$. The energy eigenvalues are equidistant and form a harmonic
spectrum truncated at $\Omega +1$ levels.  If $V_{1}$ or $V_{2}$ are
different from zero, then the energy spectrum is not harmonic any more
and only if $V_{2}\neq 0$ the energy eigenvectors correspond to a
mixing of $\left| n\right\rangle $ states.  In fact, $V_{1}K_{+}K_{-}$
is diagonal in the $\left| n\right\rangle $ space.  This term of
$ph-p^{\prime }h^{\prime }$ type does not mix states with different
particle-hole numbers.  On the contrary, the $pp^{\prime }-hh^{\prime
}$ type term, $V_{2}\left( K_{+}^{2}+K_{-}^{2}\right)$ mixes states
with $n$ and $n \pm 2$ particle-holes.  To conclude this part, we
would like to recall that the LMG model is exactly solvable using
group techniques and that it includes the Pauli exclusion principle.

\subsection{An extension of the LMG model}
The LMG model in its original form already includes some
anharmonicities, essentially those related to the fact that the Pauli
principle is treated exactly. However, parts of the residual
interaction are neglected, namely the $pp^{\prime }-p^{\prime \prime
}p^{\prime \prime \prime }$, $pp^{\prime }-p^{\prime \prime }h$,
$hh^{\prime }-h^{\prime \prime }p$, $hh^{\prime }-h^{\prime \prime
}h^{\prime \prime \prime }$ ones, which were considered in the
microscopic calculations \cite{LACCV} where the anharmonicities
arising from them were also studied.  In order to simulate them, we
propose an extension of the LMG hamiltonian which is still quadratic
in the $K_{+}$ , $K_{-}$ and $K_{0}$ operators :

\begin{equation}
H=H_{LMG}+\Delta V=H_{LMG}+V_{3}(K_{+}\bar{K}_{0}+\bar{K}_{0}K_{-})
+V_{4}(\bar{K}_{0}-1)\bar{K}_{0}
\label{h1}
\end{equation}
The $K_{+}$ $\bar{K}_{0}$ term and its hermitian conjugate introduce a
coupling between $\left| n\right\rangle $ and $\left| n\pm
1\right\rangle $ whereas the last term shifts the energies of the
$\left| n\right\rangle $ states, except those with $n=0$ and 1.
Therefore, the eigenstates $|\phi_\alpha>$ of the hamiltonian
(\ref{h1}) are superpositions of the $|n>$ states

\begin{equation}\label{phialpha}
|\phi_\alpha> = \sum_{n} X^{\alpha}_n  | n >
\end{equation}

In section 4 we will compare the exact eigenvalues of the hamiltonian
(\ref{h1}) with those corresponding to the bosonic hamiltonian
obtained by mapping the fermionic one up to the fourth order.

\section{The boson hamiltonian}
We apply boson expansion methods to the fermionic hamiltonian
(\ref{h1}) and truncated the so obtained boson hamiltonian to second
and fourth order.  In ref.s\cite{Pang,Beau} a similar study was done
for the original LMG hamiltonian (\ref{LMG}).  In the present case, the
quadratic hamiltonian corresponds to RPA while the quartic can be
directly compared with what was done in \cite{LACCV}, where some
further approximations were performed, as discussed below.

\noindent Let's take the normal ordered Holstein-Primakoff boson
expansion of the SU(2) generators \cite{Pang,Hol_Pri,Marshalek} :
\begin{equation}
\begin{array}{l}
(K_{+})_{b}=\sum_{i=0}^{\infty }a_{i}(b^{\dagger })^{i+1}b^{i} \\ 
(K_{-})_{b}=\sum_{i=0}^{\infty }a_{i}(b^{\dagger })^{i}b^{i+1} \\ 
(\bar{K}_0)_b = b^{\dagger }b
\end{array}
\label{be}
\end{equation}
where
\begin{equation}
a_{i}=\sum_{m=0}^{i}\frac{(-1)^{i-m}}{m!(i-m)!}(\Omega -m)^{1/2}
\label{ce}
\end{equation}
\noindent In the case of SU(2), the Hage-Hassan-Lambert mapping used
in ref.\cite{LACCV} is equivalent to the Holstein-Primakoff mapping
(\ref{be},\ref{ce}) \cite{Lambert} with the shifted operator
$(\bar{K}_0)_b$ .  By this shift one eliminates terms linear in
$b^{\dagger}$ and $b$ in the hamiltonian which corresponds to a
redefinition of the mean field.

\noindent Let us then consider the bosonic mapping of our extended
hamiltonian (\ref{h1}) in the particular case $V_2=V_1/2$ which is the
value used in the numerical applications we present in the next
section.  By including all the terms of the expansion (\ref{be}) which
contribute up to the second order in the $b^{\dagger},b$ operators, we
get the quadratic hamiltonian
\begin{equation}
H_{b}^{(2)}=\varepsilon (1+\delta _{1})b^{\dagger }b+\frac{\delta
_{1}\varepsilon }{2}\left( 1-\frac{1}{\Omega }\right) ^{1/2}\left(
b^{\dagger 2}+b^{2}\right)  \label{Hquad}
\end{equation}
where
\be\label{e:delta1}
\delta _{1}=\Omega V_{1}/\varepsilon
\ee
is related to the strength of the particle-hole residual interaction.
The hamiltonian $H_{b}^{(2)}$ can be written in diagonal form by
introducing the Bogoliubov transformation \cite{RS} :
\begin{equation}
\begin{array}{l}
b^{\dagger }=X Q^{\dagger }+Y Q \\ 
b=X Q+Y Q^{\dagger }
\end{array}
\label{bog}
\end{equation}
and imposing
\begin{equation}\label{20} 
H_{20}=0
\end{equation}
with the condition
\begin{equation}\label{21}  
X ^{2}-Y ^{2}=1 
\end{equation}
which guarantees the correct commutation relation 
\be\label{22}  
\left[ Q,Q^{\dagger }\right] =1
\ee
Thus one gets the RPA hamiltonian 
\begin{equation}
H_{b}^{(2)}=H_{11}Q^{\dagger }Q  \label{Hquadd}
\end{equation}
where 
\be\label{23} 
H_{11}=\varepsilon (1+\delta _{1})(X ^{2}+Y
^{2})+2X Y \delta _{1}\varepsilon \left( 1-\frac{1}{%
\Omega }\right) ^{1/2}
\ee
The conditions (\ref{20},\ref{21}) determine the $X$ and $Y$'s
amplitudes as solutions of the set of equations :
\begin{equation}
\left\{ 
\begin{array}{rcl}        
X ^{2}-Y ^{2} & = & 1 \\ 
\varepsilon (1+\delta _{1})X Y +\frac{\delta
_{1}\varepsilon }{2}\left( 1-\frac{1}{\Omega }\right)
^{1/2}(X ^{2}+Y ^{2})& = & 0\nonumber
\end{array}
\right.  \label{e:cond_bog_1}
\end{equation}
\noindent Of course the terms in $V_{3}$ and $V_{4}$ do not appear at
the quadratic order. In eq.(15), the $H_{00}$ which corresponds to a
shift of the eigenenergies was omitted.

\noindent The violations to the Pauli exclusion principle introduced
by truncating the boson expansion at the lowest order as well as a
contribution coming from the residual interaction of the $pp^{\prime
}-p^{\prime \prime }p^{\prime \prime \prime }$, $pp^{\prime
}-p^{\prime\prime }h$, $hh^{\prime }-h^{\prime \prime }p$, $hh^{\prime
}-h^{\prime\prime }h^{\prime \prime \prime }$ type can be partly
introduced by going one step further, i.e. including all terms up to
the fourth order, as it was done in ref.\cite{LACCV}.  At this level,
the terms in $V_{3}$ and $V_{4}$ will therefore enter.  Their
presence, together with the corrections for the Pauli principle, will
give rise to an anharmonic bosonic hamiltonian.

\noindent By mapping the hamiltonian of eq.(\ref{h1}) up to fourth
order we get:

\begin{equation}
\begin{array}{l}
H_{b}^{(4)}=\varepsilon (1+\delta _{1})b^{\dagger }b+\frac{\delta
_{1}\varepsilon }{2}\left( 1-\frac{1}{\Omega }\right) ^{1/2}\left(
b^{\dagger 2}+b^{2}\right) \\ 
+V_{3}\sqrt{\Omega }\left( 1-\frac{1}{\Omega }\right)^{1/2}  
\left(b^{\dagger 2}b+b^{\dagger }b^{2}\right)+ \left(V_{4}-V_{1}
\right) b^{\dagger 2}b^{2} \\ 
+\frac{\delta _{1}\varepsilon }{2}\left( 1-\frac{1}{\Omega }%
\right) ^{1/2}\left[ \left( 1-\frac{2}{\Omega }\right)^{1/2} -1\right] 
\left(b^{\dagger 3}b+b^{\dagger }b^{3}\right)
\end{array}
\label{Hquar}
\end{equation}
which we rewrite in terms of the $Q^{\dagger }$ and $Q$  operators as :
\begin{equation}
\begin{array}{r}
H_{b}^{(4)}=H_{11}Q^{\dagger }Q+H_{30}(Q^{\dagger
3}+Q^{3})+H_{21}(Q^{\dagger 2}Q+Q^{\dagger }Q^{2})+ \\ 
H_{31}(Q^{\dagger 3}Q+Q^{\dagger }Q^{3})+H_{22}Q^{\dagger
2}Q^{2}+H_{40}(Q^{\dagger 4}+Q^{4})
\end{array}
\label{Hquarr}
\end{equation}
whose coefficients are given in the appendix
(eqs.\ref{e:H_11}-\ref{e:L}), together with the new equations for $X$
and $Y$ (\ref{e:cond_bog_2}) coming again from the conditions
(\ref{20},\ref{21}).  In eq.(\ref{Hquarr}) the $H_{00}$ was omitted as
well as a term, linear in the $Q^{\dagger}$ and $Q$ operators, which
would introduce a redefinition of the mean field, in analogy with what
was done in \cite{LACCV}.

\noindent The quartic hamiltonian (\ref{Hquarr}) corresponds to that
used in \cite{LACCV} where, however, some approximations were
introduced in order to make feasible the calculations in the realistic
cases considered there.  First, only one- and two-phonon states were
considered.  Therefore, the terms in $H_{30}$, $H_{31}$ and $H_{40}$
were not effective.  Second, the $X$ and $Y$ amplitudes appearing in
the fourth order hamiltonian were not recalculated but taken equal to
those obtained at the second order, i.e. the RPA ones. In order to get
an indication on how good these approximations on the space, on the
hamiltonian and on the $X$ and $Y$ amplitudes are, in the next section
we will study them within the present schematic model.

\section {Results and discussion}
First of all we have to fix the parameters entering in the hamiltonian
(\ref{h1}). For the single particle energy we use the parametrization
$\varepsilon = 41/A^{1/3}$ MeV. We take $V_2 = 0.5 V_1$ and the
strength $V_1$ is fixed so that the first excited state lies at an
energy around $80/A^{1/3}$ MeV corresponding to the systematics of GDR
in nuclei.  This criterion gives $V_1=1.2$~MeV.  We have studied the
behaviour of the energies of the three lowest states, $E_1$, $E_2$ and
$E_3$, as a function of $V_4$ for two values of $V_3$, namely
$V_3=0$~MeV (Fig.1) and $V_3=0.25$~MeV (Fig.2).  The latter value for
$V_3$ gives $<2|\Delta V|1> \approx 1$~MeV in analogy with the microscopic
calculations \cite{LACCV}.  Note that the sign of $V_3$ is
irrelevant. As far as the sign of $V_4$ is concerned, we show results
only for negative values, which give a downward shift of $E_2$ and
$E_3$ with respect to $2E_1$ and $3E_1$, respectively, i.e. the
harmonic (RPA) values. In the figures we show the results obtained for
$\Omega=8$.

\begin{figure}[htb]
\mbox{\epsfig{file=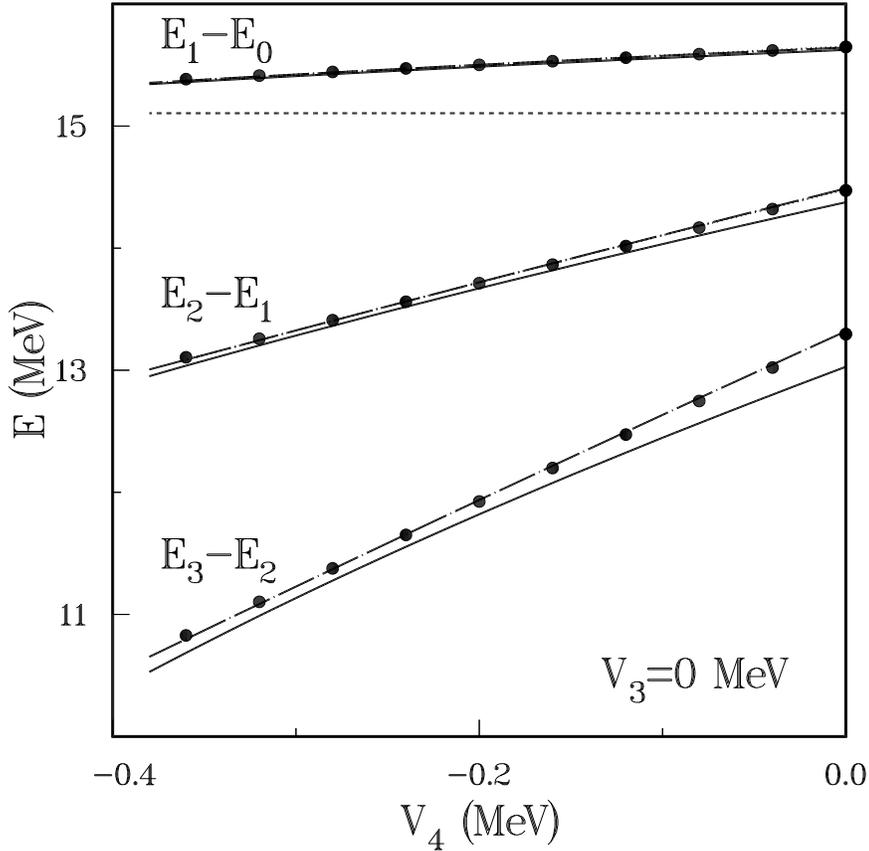,height=13.0truecm,angle=0}}
\caption{ Energy differences as functions of $V_4$ for a
fixed value of $V_3=0$ MeV. The ones corresponding to the fermionic
hamiltonian (\ref{h1}) are plotted as solid lines. The others
correspond to different approximations on the bosonic expansion of the
hamiltonian (see text).}
\end{figure}
\hspace{\fill}
\begin{figure}[htb]
\mbox{\epsfig{file=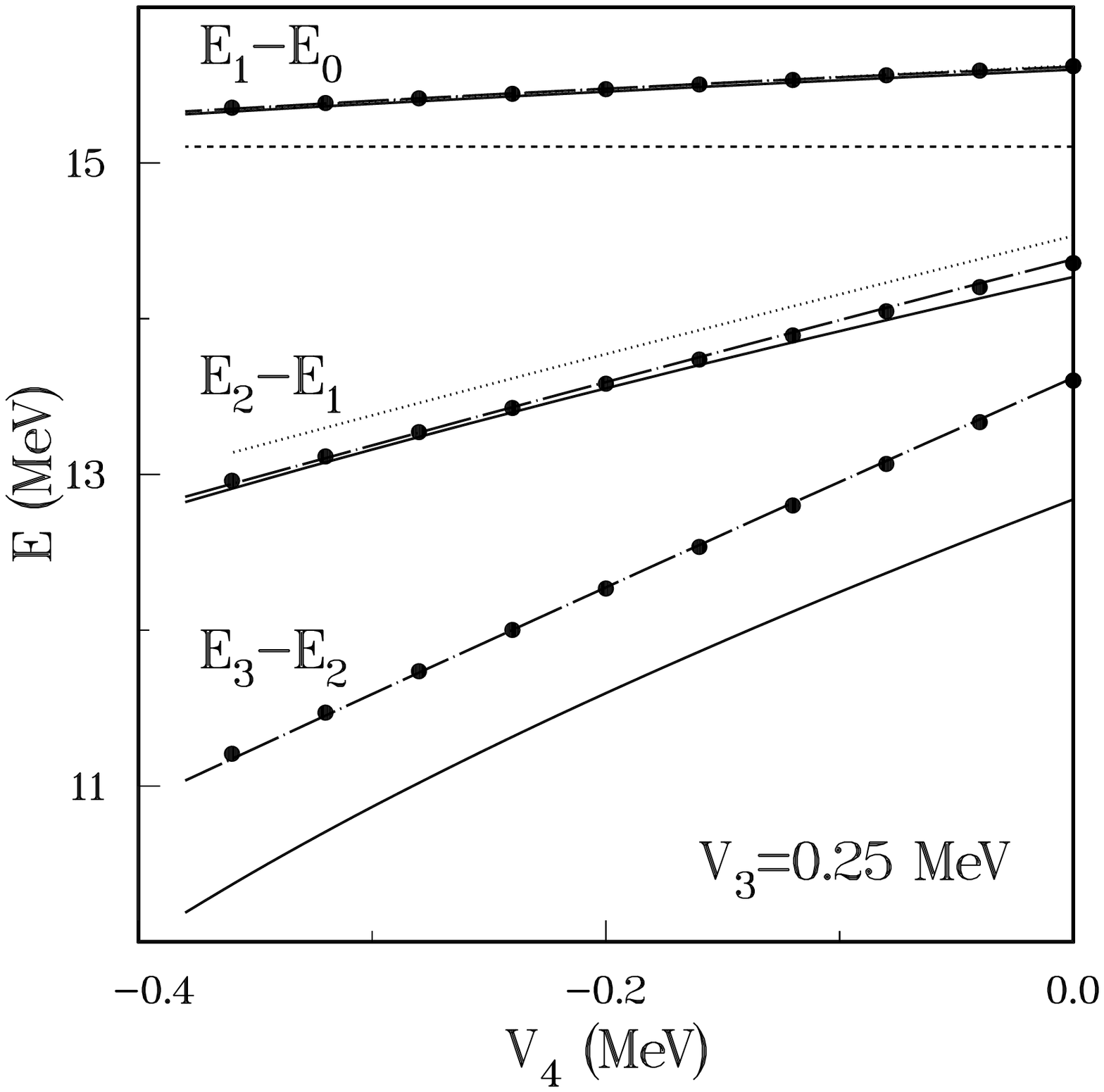,height=13.0cm,angle=0}}
\caption{As in fig. 1, but for $V_3=0.25$ MeV.}
\end{figure}

\noindent To compare the spectrum of the exact hamiltonian and its
different boson expansions we study the energy differences $(E_{n} -
E_{n-1})$ between the lower excited states. In RPA they have the
common value 15.11 MeV for the considered parameters, independent of
n, $V_{3}$ and $V_{4}$. In the figures this value is represented by a
dashed line. This should be compared with the exact results for the
fermionic hamiltonian eq.(5), shown as solid lines. We see that, even
for $V_{3} = V_{4} = 0$, the RPA results deviate from the exact
ones. Namely, the first three differences obtained in the exact
calculations are equal to 15.62 MeV, 14.38 MeV and 13.03 MeV,
respectively.

\noindent The agreement with the exact energies can be improved by
using the quartic hamiltonian, eq.(19).  When the bosonic hamiltonian
is diagonalized in the space containing up to two-phonon states we get
the results presented in Fig. 1 and 2 by dotted lines. We remark that
now the first excited state is very close to the exact one, while for
the second one there is a discrepancy of not more than 250 KeV. In the
more realistic microscopic calculations presented in \cite{LACCV} a
further approximation was introduced. Instead of reobtaining the $X$
and $Y$ amplitudes by solving the equations analogous to eqs (20), the
RPA amplitudes were used. In the present model, this corresponds to
use in eq. (19) the solutions of eqs (17) rather than those of
(20). The results obtained are indistinguishable from the dotted lines
in Fig. 1 and 2.  This supports the validity of the procedure used in
\cite{LACCV}.

\noindent In order to test how much the truncation of the space
affects the results, let us now enlarge the space up to three-phonon
states.  Of course, in this space, the quartic term $H_{40}$ in (19)
does not play any role.  In the figures, we show the results
corresponding to the complete quartic hamiltonian of
eq.(\ref{Hquarr}), with the $X$ and $Y$ amplitudes solutions of eq.s
(20) (dot-dashed lines).  Comparing the energies of the first two
states we see that the agreement with the exact ones is now almost
perfect. In the enlarged space we can also study the third excited
state which, in the harmonic limit, would correspond to a three-phonon
state. In this case the approximate results differ from the exact ones
by 1 MeV, which can be considered as a good approximation since the
excitation energy of this state is about 40 MeV. We would like to
point out that the results obtained in the same model space but
corresponding to the quartic hamiltonian of eq (19) without the
$H_{31}$ and $H_{30}$ term and with the use of the RPA $X$ and $Y$
amplitudes, solution of eqs (17), (represented in figure 1 and 2 like
solid circles) are practically coincident with the dot-dashed line. The use 
of an analogous approximation in realistic microscopic calculations would make
them much simpler.

\noindent In summary, we have seen that quite good results can be obtained 
for both the second and the third excited state by diagonalizing the fourth 
order bosonic hamiltonian with coefficients calculated with the RPA $X$ and 
$Y$ amplitudes, in the space containing up to three-phonon states. Nowadays 
there is interest in searching for triple giant resonance excitations 
\cite{triple}. The present results indicate that a realistic theoretical 
study of such states will be feasible in the near future.

\section{Conclusions}
We have applied boson expansion methods to an extended
Lipkin-Meshkov-Glick model including a residual interaction between
the excited states in analogy with the anharmonicities introduced in
the microscopic calculations of ref.\cite{LACCV}. We have studied the
effect of truncations of the space and of the hamiltonian as well as
other approximations present in those calculations by comparing the
approximate results on the energies of the first three excites states
among them and with the exact ones.

\noindent From the analysis presented we can conclude that the
approximations made in \cite{LACCV} are well justified if one wants to
study one- and two-phonon excitations. What about three-phonon states?
Of course, in the context of boson expansion methods, the study of
these states would need a boson expansion up to sixth order, but this
would be a formidable task.  The results shown make us confident that
a quartic hamiltonian as that used in \cite{LACCV}, diagonalized in an
enlarged space including up to three-phonon states, may give
reasonable results also in realistic cases.

\vspace{5mm} Two of the authors (F.C. and E.G.L.) are grateful, for the
warm hospitality, to the Departamento de FAMN of the University of
Sevilla where part of the work has been done. E.G.L. is a Marie Curie
Fellow with contract n. ERBFMBICT983090 within the TMR programme of
the Europen Community.  This work has been partially supported by the
spanish DGICyT under contract PB95-0533-A, by the Spanish-Italian
agreement between the DGICyT and the INFN and by the Spanish-French
agreement between the DGICyT and the IN2P3.

\section{Appendix}

The $X$ and $Y$ coefficients of the Bogoliubov transformation obtained
by imposing conditions (\ref{20},\ref{23}) for the quartic expansion
$H^{(4)}_{b}$ of the hamiltonian (\ref{h1}) are solutions of the
equations
\begin{equation}  \label{e:cond_bog_2}
\left \{ 
\begin{array}{rcl}
X ^{2}-Y ^{2} & = & 1 \\ 
\varepsilon (1+\delta _{1})X Y +\frac{\delta _{1}\varepsilon}{2}
\left( 1-\frac{1}{\Omega }\right) ^{1/2}(X ^{2}+Y ^{2})+
&  &  \\ 
\left( V_{4}-V_{1}\right) X Y \left( X ^{2}+5Y
^{2}\right) +3\lambda \left( 3X ^{2}Y ^{2}+Y ^{4}\right) & = & 
0 \nonumber
\end{array}
\right.
\end{equation}

The coefficients of the boson hamiltonian $H^{(4)}_{b}$ are given by
the following expressions

\begin{equation}
\begin{array}{l}
H_{11}=\varepsilon (1+\delta _{1})(X ^{2}+Y ^{2})+2X
Y \delta _{1}\varepsilon \left( 1-\frac{1}{\Omega }\right) ^{1/2}+\\
4\left(2X ^{2}Y ^{2}+Y ^{4}\right)\left(V_{4}-V_{1}\right) +
6X Y\lambda \left(X^{2}+3Y ^{2}\right)
\end{array}
\label{e:H_11}
\end{equation}

\noindent

\begin{equation}
H_{30}=V_{3}X Y \left( X +Y \right) \sqrt{\Omega }\left( 1-%
\frac{1}{\Omega }\right) ^{1/2}  \label{e:H_30}
\end{equation}

\begin{equation}
H_{21}=V_{3}\sqrt{\Omega }\left( 1-\frac{1}{\Omega }\right) ^{1/2}\left(
X ^{3}+2X Y ^{2}+2X ^{2}Y +Y ^{3}\right)
\label{e:H_21}
\end{equation}

\begin{equation}
H_{40}=\left( -V_{1}+V_{4}\right) X ^{2}Y ^{2}+\lambda X Y
(X ^{2}+Y ^{2})  \label{e:H_40}
\end{equation}

\begin{equation}
H_{31}=\left( -V_{1}+V_{4}\right) 2X Y \left( X^2 +Y^2 \right)
+\lambda \left( X ^{4}+6 X^2 Y^2 + Y ^{4}\right)
\label{e:H_31}
\end{equation}

\begin{equation}
H_{22}=\left( -V_{1}+V_{4}\right) \left( X ^{4}+4X ^{2}Y
^{2}+Y ^{4}\right) +6X Y \lambda \left( X ^{2}+Y
^{2}\right)  \label{e:H_22}
\end{equation}

and

\begin{equation}
\lambda =\frac{\delta _{1}\varepsilon }{2}\left( 1-\frac{1}{\Omega 
}\right) ^{1/2}\left[ \left( 1-\frac{2}{\Omega }\right) ^{1/2}-1\right]
\label{e:L}
\end{equation}
In particular, if $V_{1}=V_{3}=V_{4}=0$, the relations
(\ref{e:H_11})-(\ref{e:L}) coincide with those of reference \cite{Pang}.

\end{document}